\documentclass[twocolumn]{aastex631}

\submitjournal{\apjl}

\begin{document}

\title{The Luminosity Function of Tidal Disruption Flares for the ZTF-I Survey}

\author[0000-0003-4959-1625]{Zheyu Lin}
\affiliation{Deep Space Exploration Laboratory / Department of Astronomy, University of Science and Technology of China, Hefei 230026, China; linzheyu@mail.ustc.edu.cn, jnac@ustc.edu.cn, xkong@ustc.edu.cn}
\affiliation{School of Astronomy and Space Sciences, 
University of Science and Technology of China, Hefei, 230026, China}

\author[0000-0002-7152-3621]{Ning Jiang}
\affiliation{Deep Space Exploration Laboratory / Department of Astronomy, University of Science and Technology of China, Hefei 230026, China; linzheyu@mail.ustc.edu.cn, jnac@ustc.edu.cn, xkong@ustc.edu.cn}
\affiliation{School of Astronomy and Space Sciences, 
University of Science and Technology of China, Hefei, 230026, China}
\affiliation{Frontiers Science Center for Planetary Exploration and Emerging Technologies, University of Science and Technology of China, Hefei, Anhui, 230026, China}

\author[0000-0002-7660-2273]{Xu Kong}
\affiliation{Deep Space Exploration Laboratory / Department of Astronomy, University of Science and Technology of China, Hefei 230026, China; linzheyu@mail.ustc.edu.cn, jnac@ustc.edu.cn, xkong@ustc.edu.cn}
\affiliation{School of Astronomy and Space Sciences, 
University of Science and Technology of China, Hefei, 230026, China}
\affiliation{Frontiers Science Center for Planetary Exploration and Emerging Technologies, University of Science and Technology of China, Hefei, Anhui, 230026, China}

\author[0000-0001-7689-6382]{Shifeng Huang}
\affiliation{Deep Space Exploration Laboratory / Department of Astronomy, University of Science and Technology of China, Hefei 230026, China; linzheyu@mail.ustc.edu.cn, jnac@ustc.edu.cn, xkong@ustc.edu.cn}
\affiliation{School of Astronomy and Space Sciences, 
University of Science and Technology of China, Hefei, 230026, China}

\author[0000-0001-8078-3428]{Zesen Lin}
\affiliation{Deep Space Exploration Laboratory / Department of Astronomy, University of Science and Technology of China, Hefei 230026, China; linzheyu@mail.ustc.edu.cn, jnac@ustc.edu.cn, xkong@ustc.edu.cn}
\affiliation{School of Astronomy and Space Sciences, 
University of Science and Technology of China, Hefei, 230026, China}

\author[0000-0003-3824-9496]{Jiazheng Zhu}
\affiliation{Deep Space Exploration Laboratory / Department of Astronomy, University of Science and Technology of China, Hefei 230026, China; linzheyu@mail.ustc.edu.cn, jnac@ustc.edu.cn, xkong@ustc.edu.cn}
\affiliation{School of Astronomy and Space Sciences, 
University of Science and Technology of China, Hefei, 230026, China}

\author[0000-0003-4225-5442]{Yibo~Wang}
\affiliation{Deep Space Exploration Laboratory / Department of Astronomy, University of Science and Technology of China, Hefei 230026, China; linzheyu@mail.ustc.edu.cn, jnac@ustc.edu.cn, xkong@ustc.edu.cn}
\affiliation{School of Astronomy and Space Sciences, 
University of Science and Technology of China, Hefei, 230026, China}



\begin{abstract}
The high-cadence survey of Zwicky Transient Facility (ZTF) has completely dominated the discovery of tidal disruption events (TDEs) in the past few years and resulted in the largest sample of TDEs with optical/UV light curves well-sampled around their peaks, providing us an excellent opportunity to construct a peak luminosity function (LF) of tidal disruption flares (TDFs).
The new construction is necessary particularly considering that the most updated LF reported in literature has been inferred from only 13 sources from 5 different surveys.
Here we present the optical and blackbody LFs calculated by 33 TDFs discovered in the ZTF-I survey.
The optical LF can be described by both a power-law profile $dN/dL_g\propto L_g^{-2.3\pm0.2}$, and a Schechter-like function. The blackbody LF can be described by a power-law profile $dN/dL_{\rm bb}\propto L_{\rm bb}^{-2.2\pm0.2}$, shallower than the LF made of previous \citet{vanVelzen18} sample. 
A possible low-luminosity turnover in the optical LF supports an Eddington-limited emission scenario. The drop of volumetric rate at high luminosity suggests a rate suppression due to direct captures of the black hole. The total volumetric rate is one order of magnitude lower than the previous estimation, which is probably not simply caused by the high fraction post-peak sources (7/13) in the previous sample. Instead, the normalization step during the previous LF construction to reconcile various surveys might adversely amplify the influence of serendipitous discoveries. 
Therefore, TDFs selected from ongoing and upcoming uniform surveys like ZTF, Vera Rubin Observatory (VRO) and Wide-Field Survey Telescope (WFST) should yield more accurate LFs.


\end{abstract}



\section{Introduction} \label{sec:intro}
A tidal disruption event (TDE) happens when a massive black hole (BH) in the galaxy center tears apart an unlucky star that wanders too close, as the tidal forces overcome the self-gravity of the star. About half of the stellar debris will circularize and be accreted by the BH, producing a luminous flare that peaks from ultraviolet (UV) to soft X-ray band \citep[e.g.,][]{Rees88,Gezari21}. The occurrence rate of TDEs is about $10^{-4}-10^{-5}$ galaxy$^{-1}$ year$^{-1}$ \citep[e.g.,][]{Wang04,Stone16}. Although the first TDE was discovered in the X-ray band \citep{Bade96}, the optical bands have dominated the discovery of TDEs in the past decade, thanks to the large survey area and the high cadence of the time-domain surveys, such as All-Sky Automated Survey for Supernovae \citep[ASAS-SN,][]{Shappee14} and the Zwicky Transient Facility \citep[ZTF,][]{Bellm19a}. The total number of TDEs is now increasing at the rate of $\sim10-20$ per year, and is predicted to skyrocket at the pace of a few thousand per year, as the larger surveys by Vera Rubin Observatory \citep[VRO,][]{Ivezic19} and the Wide-Field Survey Telescope (WFST) are both scheduled to begin in 2023 \citep[e.g.,][]{Strubbe09,vanVelzen11,Thorp19,Bricman20,Roth21,Lin22}.


Tidal disruption flares (TDFs) in optical bands are mainly characterized by three features. First, the spectral energy distribution (SED) can be well described by a blackbody component, with the temperature varies slowly between $(1-5)\times10^4$ K. This results in the steady blue color in optical bands (e.g. $g-r<0$). Second, the optical light curve of a TDF usually shows a monthly rise to a peak blackbody luminosity of $L_{\rm{bb}}\sim10^{43-45}$ erg s$^{-1}$, followed by a power-law like decline that lasts for months to years. Third, the broad H$\alpha$, H$\beta$ or He $\textsc{ii}$ emission lines usually appear in the spectra of TDFs near the peak, and then gradually narrow and weaken as the luminosity declines. Nonetheless, emission line features in some TDFs are too intense to distinguished, as a strong blue continuum dominates their spectra. These features are extracted from previous serendipitous discoveries, and now they instruct the search of TDFs for some large transient surveys. Most outstandingly, the ZTF-I survey has discovered around 30 TDFs, which consists the largest systematical sample of TDFs \citep{Hammerstein22}. 

In spite of a persistent number increase and gradually improved characterization of optical TDEs, their real event rate and emission origin remain rather uncertain~\citep[e.g.,][]{Piran15,Jiang16,Metzger16,Dai18,Lu20}. The TDE event rate can indicate the stellar dynamics in galaxy centers, which determines how stars are fed to the loss cone \citep[e.g.,][]{Merritt04,Stone20}. Earlier rough TDE rate estimation based on very few luminous events, such as the two in SDSS~\citep{vV2014} and the other two in ASAS-SN survey~\citep{Holoien2016}, resulted in a significantly lower rate than theoretical expectation albeit with large uncertainties. A well-constrained luminosity function (LF) extended to faint end can definitely help make progress on the measurement.
Meanwhile, the LF might also shed useful insight into the physical process following the disruption and the correlation between optically and X-ray bright TDEs from a statistical point of view.

However, the most updated LF of optical/UV TDF remains the one proposed by \citet{vanVelzen18}, which is constructed with a sample of 13 sources, that is almost all optical/UV TDFs up to then. It is worthwhile to note that 7 out of 13 sources have only post-peak light curves, indicating a possible underestimate of peak luminosity. As a result, the maximum redshift ($z_{\rm{max}}$) for detection and identification of these sources might be underestimated and the volumetric rate of TDFs is thus overestimated. In addition, these TDFs are discovered by 5 surveys ($GALEX$, SDSS, PS1, iPTF, ASAS-SN) with different survey areas, depths, filters and cadences. Therefore, a normalization based on the TDF number in each survey has to be applied. This may amplify the influence for some serendipitous searches. In contrast, the TDF sample given by ZTF-I survey has several obvious advantages. First, it is $\sim2.5$ times larger than the previous sample. Second, almost all TDFs are captured before or around their peaks, allowing a reliable measurement of their peak luminosity. Third, the TDFs are systematically selected by a single detector, precluding the serendipity. Last but not least, most TDFs are also followed by $Swift$ UVOT in the UV bands, which is very useful to constrain their blackbody temperature, radius and luminosity. In a word, it is meaningful to reconstruct a LF using the TDF sample of ZTF-I survey. 

The outline of the paper is as follows. We will first introduce the TDF sample of ZTF-I survey, the follow-up observations of $Swift$ UVOT and the additional photometric data in Section \ref{sec:sampledata}. Next, in Section \ref{sec:bbfit}, we will describe the method of fitting to a blackbody model, and show the fitting results. We will construct the LF in Section \ref{sec:lf}, then briefly discuss the results in Section \ref{sec:dis}. Finally, we conclude in Section \ref{sec:con}. We adopt a flat cosmology with $H_{0} =70$ km~s$^{-1}$~Mpc$^{-1}$ and $\Omega_{\Lambda} = 0.7$. All magnitudes are in the AB \citep{Oke74} system.

\subsection{TDE or TDF}
In the literature, both TDE and TDF are used to label transients due to stellar disruptions. In this work, TDE refers to a class of events, in which a BH disrupts a star, while TDF represents for the electromagnetic emission that can be classified as a TDE. This distinction is subtle, yet useful, since some TDEs may not lead to TDFs, due to delayed accretion \citep[e.g.,][]{Guillochon15}, or prohibited accretion \citep[e.g.,][]{Bonnerot16,Coughlin17}.

\section{TDF Sample \& Data} \label{sec:sampledata}
\subsection{ZTF-I Survey}\label{sec:ztfi}
ZTF-I survey is the first phase of the ZTF survey. It started in March 2018 and completed in October 2020. During the 2.7-year survey, a systematic search for TDEs had been conducted almost entirely with the public MSIP data \citep{Bellm19b}, which observed the entire visible Northern sky every 3 nights in both $g$- and $r$-bands. \citet{Hammerstein22} has reported a sample of 30 photometrically and spectroscopically classified TDFs.
We add all these 30 TDFs into our sample. In addition, we check the Transient Name Server\footnote{\url{https://www.wis-tns.org}} (TNS) and find 4 more spectroscopically classified TDFs, that were also reported by ZTF during the ZTF-I survey. These sources are AT2019gte, AT2020neh, AT2020nov and AT2020vwl.

We perform forced point spread function (PSF) photometry to extract precise flux measurements of each source through the ZTF forced-photometry service \citep{Masci19}. We clean the photometry results by filtering out epochs that are impacted by bad pixels, and requiring thresholds for the signal-to-noise of the observations, seeing, zeropoint, the sigma-per-pixel in the input science image, and the 1-$\sigma$ uncertainty on the difference image photometry measurement. We perform baseline corrections for sources whose differential fluxes are significantly offset from zero counts (0 DN) in the quiescent state. For each of these sources, we first classify the measurements by the field, charge-coupled device (CCD) and quadrant identifiers. Then, for each class, we set the median of pre- or sufficiently post-flare counts as the offset. These sources are AT2018zr, AT2018bsi, AT2018hco, AT2018hyz, AT2018lna, AT2019dsg, AT2019qiz, AT2019meg, AT2020ocn and AT2020qhs.

After the initial filtering, we construct the differential light curves for all 34 sources, including corrections for the Galactic extinction. We adopt the \citet{Planck16} GNILC dust map, and use the {\tt dustmaps} package \citep{Green18} to evaluate the Galactic extinction. 
We have carefully checked the light curves of the four TNS sources, and exclude AT2019gte from the final sample since it shows a quick color evolution from blue to red around the peak, which is distinct from other TDFs. We thus obtain a final sample of 33 TDFs selected from ZTF-I survey. 


\subsection{Swift UVOT Observations}\label{sec:uvot}
All 33 TDFs were followed up with observations from the $Neil\ Gehrels\ Swift\ Observatory$ \citep{Gehrels04} in the UV with UVOT \citep{Roming05}. We use the {\tt uvotsource} package to perform $Swift$ UVOT photometry. In order to capture the light of the TDFs and their host galaxies, we carefully examine the image files for each source, and use different apertures of 5$\arcsec$, 7$\arcsec$, 10$\arcsec$ or 20$\arcsec$ depending on the size of the sources. 
We first build the reference images with {\tt uvotimsum} package using data that are observed late enough, then perform photometry on them. Then we subtract the reference fluxes and get the differential fluxes of $Swift$ UVOT. If the variability in some bands is not obvious 
comparing with reference images or there is no proper late-stage images for reference, we will not take these bands into consideration when we perform the blackbody model fitting, as described in Section \ref{sec:bbfit}.

\subsection{Additional Photometric Data}\label{sec:add}
Besides ZTF and $Swift$, we also obtain photometric data from the Asteroid Terrestrial-impact Last Alert System (ATLAS) survey using the ATLAS forced photometry service\footnote{\url{https://fallingstar-data.com/forcedphot/}} \citep{Tonry18,Smith20}, the open TDE catalog\footnote{\url{https://tde.space}. Unfortunately, data on this website can not be retrieved recently.} \citep{Guillochon17}, and bibliography for individual source. 

For the ATLAS forced photometry service, the positions for all sources have been observed during the flare, regardless the significance of the flares. 
We have removed the epochs that with large errors, and performed baseline corrections using the median flux pre- or sufficiently post-flare. Then we combine the data in 1-day bins. We note that the ATLAS photometry service sometimes changes the reference image for differential photometry during the flare, which causes an unrecoverable offset. In this situation we have to discard the light curves.

Only 4 sources have been collected by the open TDE catalog: AT2018zr, AT2018hco, AT2018hyz, AT2019qiz. We download the {\tt json} files that contain the photometric data, and refer to the cited papers to examine if the reference fluxes have been subtracted \citep{vanVelzen19,Nicholl20,Gomez20}. If not subtracted, we cast away the data of this band. We utilize most of the data, including observations by other optical telescopes like Las Cumbres Observatory Global Telescope (LCOGT) network 
\citep{Brown13}, Palomar 60-inch telescope (P60), Pan-STARRS \citep{Chambers16} and ASAS-SN.

\vspace{10pt}

\begin{figure*}
\plottwo{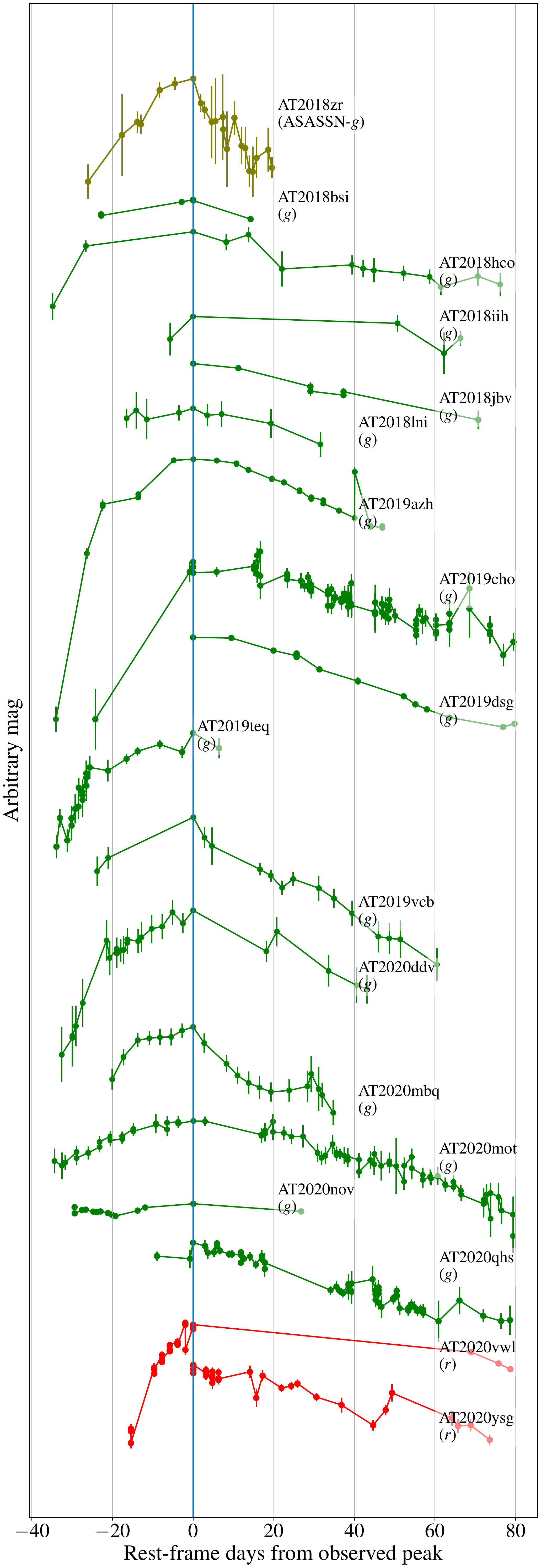}{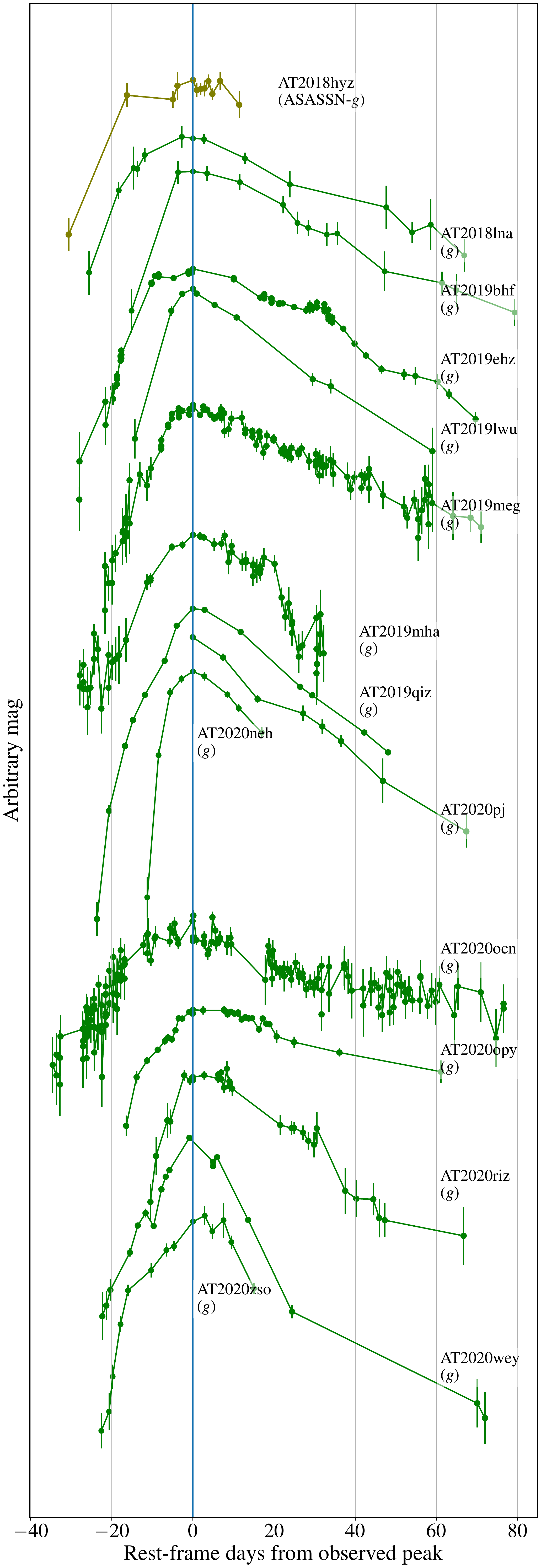}
\caption{Light curves around the peak for ZTF-I TDFs.\label{fig:lc}}
\end{figure*}

In summary, we present the entire sample of 33 TDFs selected from ZTF-I survey in Table \ref{tab:ztfi}, along with the IAU name, ZTF name, names given by other surveys, discovery date, coordinates and redshift. Light curves around the peak for all TDFs are displayed in Figure \ref{fig:lc}.

\begin{deluxetable*}{lllcccl}[!htbp]\label{tab:ztfi}
\tablecaption{ZTF-I TDF Sample}
\tablewidth{0pt}
\tablehead{
IAU Name  & ZTF Name              & Other Name(s)                    & Discovery Date & R.A.\,(J2000) & Dec.\,(J2000) & Redshift
}

\startdata
AT2018zr  & ZTF18aabtxvd          & \textbf{PS18kh}/ATLAS18nej       & 2018-03-02       & 07:56:54.54  & +34:15:43.6  & 0.075    \\
AT2018bsi & \textbf{ZTF18aahqkbt} &                                  & 2018-04-09       & 08:15:26.62  & +45:35:31.9  & 0.051    \\
AT2018hco & ZTF18abxftqm          & \textbf{ATLAS18way}              & 2018-10-04       & 01:07:33.64  & +23:28:34.3  & 0.088    \\
AT2018hyz & ZTF18acpdvos          & \textbf{ASASSN-18zj}/ATLAS18bafs & 2018-11-06       & 10:06:50.87  & +01:41:34.1  & 0.0457
\\
AT2018iih & ZTF18acaqdaa          & \textbf{ATLAS18yzs}/Gaia18dpo    & 2018-11-09       & 17:28:03.93  & +30:41:31.4  & 0.212    \\
AT2018jbv & \textbf{ZTF18acnbpmd} & ATLAS19acl/PS19aoz               & 2018-11-26       & 13:10:45.56  & +08:34:04.3  & 0.34     \\
AT2018lni & \textbf{ZTF18actaqdw} &                                  & 2018-11-28       & 04:09:37.65  & +73:53:41.6  & 0.1380   \\
AT2018lna & \textbf{ZTF19aabbnzo} &                                  & 2018-12-28       & 07:03:18.65  & +23:01:44.7  & 0.0910   \\
AT2019bhf & \textbf{ZTF19aakswrb} &                                  & 2019-02-12       & 15:09:15.97  & +16:14:22.5  & 0.1206   \\
AT2019cho & \textbf{ZTF19aakiwze} &                                  & 2019-02-12       & 12:55:09.23  & +49:31:09.8  & 0.1930   \\
AT2019azh & ZTF17aaazdba          & ASASSN-19dj/\textbf{Gaia19bvo}            & 2019-02-22       & 08:13:16.94  & +22:38:54.0  & 0.0223   \\
AT2019dsg & \textbf{ZTF19aapreis} & ATLAS19kl                        & 2019-04-09       & 20:57:02.97  & +14:12:15.9  & 0.0512   \\
AT2019ehz & ZTF19aarioci          & \textbf{Gaia19bpt}               & 2019-04-29       & 14:09:41.88  & +55:29:28.1  & 0.0740   \\
AT2019lwu & \textbf{ZTF19abidbya} & ATLAS19rnz/PS19ega               & 2019-07-24       & 23:11:12.31  & $-$01:00:10.7  & 0.117    \\
AT2019meg & \textbf{ZTF19abhhjcc} & Gaia19dhd                        & 2019-07-28       & 18:45:16.20  & +44:26:19.1  & 0.152    \\
AT2019mha & ZTF19abhejal          & \textbf{ATLAS19qqu}              & 2019-07-30       & 16:16:27.85  & +56:25:56.3  & 0.148    \\
AT2019qiz & \textbf{ZTF19abzrhgq} & ATLAS19vfr/Gaia19eks/PS19gdd     & 2019-09-19       & 04:46:37.88  & $-$10:13:34.9  & 0.0151   \\
AT2019teq & \textbf{ZTF19accmaxo} &                                  & 2019-10-20       & 18:59:05.50  & +47:31:05.7  & 0.0878   \\
AT2019vcb & \textbf{ZTF19acspeuw} & ATLAS19bcyz/Gaia19feb            & 2019-11-15       & 12:38:56.38  & +33:09:57.3  & 0.089    \\
AT2020pj  & \textbf{ZTF20aabqihu} & ATLAS20cab                       & 2020-01-02       & 15:31:34.96  & +33:05:41.5  & 0.068    \\
AT2020ddv & ZTF20aamqmfk          & \textbf{ATLAS20gee}              & 2020-02-21       & 09:58:33.37  & +46:54:40.1  & 0.16     \\
AT2020ocn & \textbf{ZTF18aakelin} &                                  & 2020-04-29       & 13:53:53.77  & +53:59:49.5  & 0.07     \\
AT2020mbq & \textbf{ZTF20abefeab} & ATLAS20pfz/PS20grv               & 2020-06-09       & 15:40:15.26  & +25:00:04.8  & 0.093    \\
AT2020mot & \textbf{ZTF20abfcszi} & Gaia20ead                        & 2020-06-14       & 00:31:13.56  & +85:00:31.9  & 0.07     \\
AT2020neh & \textbf{ZTF20abgwfek} & ATLAS20qkz/Gaia20cxg/PS20elo     & 2020-06-19       & 15:21:20.09  & +14:04:10.5  & 0.062    \\
AT2020nov & \textbf{ZTF20abisysx} & ATLAS20vfv/Gaia20duz/PS20ggg     & 2020-06-27       & 16:58:12.97  & +02:07:03.0  & 0.084    \\
AT2020opy & \textbf{ZTF20abjwvae} & PS20fxm                          & 2020-07-08       & 15:56:25.72  & +23:22:20.8  & 0.159    \\
AT2020qhs & \textbf{ZTF20abowque} & ATLAS20upw/PS20krl               & 2020-07-26       & 02:17:53.95  & $-$09:36:50.8  & 0.345    \\
AT2020riz & \textbf{ZTF20abrnwfc} & PS20jop                          & 2020-07-31       & 02:10:30.75  & +09:04:26.5  & 0.435    \\
AT2020wey & \textbf{ZTF20acitpfz} & ATLAS20belb/Gaia20fck            & 2020-10-08       & 09:05:25.88  & +61:48:09.2  & 0.0274  \\
AT2020vwl & ZTF20achpcvt          & ATLAS20bdgk/\textbf{Gaia20etp}            & 2020-10-10       & 15:30:37.80  & +26:58:56.9  & 0.035    \\
AT2020ysg & \textbf{ZTF20abnorit} & ATLAS20bjqp/PS21cru              & 2020-10-28       & 11:25:26.03  & +27:26:26.2  & 0.277    \\
AT2020zso & \textbf{ZTF20acqoiyt} & ATLAS20bfok                      & 2020-11-12       & 22:22:17.13  & $-$07:15:59.1  & 0.061   
\enddata    
\tablecomments{The names of each TDE detected in ZTF-I, with boldface indicating the discovery name, i.e. the first survey to report photometry of the transient detection to the TNS.}
\end{deluxetable*}

\section{Blackbody Fitting} \label{sec:bbfit}
After collecting the photometric data of the 33 TDFs, we begin to fit the peak SED of each source into a blackbody model. We use a simplest redshifted blackbody model, which only contains two free parameters: the temperature $T_{\rm bb}$ and radius $R_{\rm bb}$,
\begin{equation}
    f_{\lambda}(\lambda)=\frac{2\pi hc^2}{\lambda^5}(1+z)^4
    \frac{1}{e^{hc(1+z)/\lambda kT_{\rm bb}}-1}\left(\frac{R_{\rm bb}}{D_L(z)}\right)^2.
\end{equation}

Before the fitting, we determine the peaks by ZTF light curves except for AT2018zr, 
which peaked during the phase of reference image construction, and consequently the ZTF forced-photometry service can not provide the differential light curves around the peak. For this source we refer to \citet{vanVelzen19}, in which the peak was carefully calibrated after the reconstruction of reference images. Then we extract the optical photometric data around the peak for each source.

To constrain the blackbody temperature more accurately, we also add the $Swift$ UVOT data into the SED fitting especially taking advantage of its unique UV photometry. As mentioned in Section \ref{sec:uvot}, we have only selected UVOT bands with both late-time observations and significant variability to ensure a reliable measurement. These two selection criteria filter out sources that are not extensively covered by $Swift$ and result in 20 sources being followed sufficiently well in at least one UVOT band.
Among them, 11 sources have UVOT coverage around ZTF peaks and the other 9 have only available observations in the declining phase.
For the latter case, we assume that the $T_{\rm bb}$ evolves slowly, which is a common feature of TDE, hence the variation since peak should be negligible. The assumption is reasonable for most sources that were observed by UVOT $\sim10-50$ days after their peaks.
We fit these post-peak data, and use the resulting $T_{\rm bb}$ and the peak data to calculate the peak $R_{\rm bb}$.

For the remaining 13 sources, UV photometry is unfortunately absent.
In order to minimize the impact of too few bands, we have taken ZTF $i$ band, ATLAS $c$ and $o$ bands and some LCOGT and ASAS-SN bands into consideration. However, the fitting to most sources still result in large uncertainties. We note that AT2020ysg is the only source whose peak can not be resolved, due to a large $\sim80$-day gap around its intrinsic peak. In addition, its detected peak has only a single measurement, hence a post-peak fitting is still needed.



To get the two parameters, $T_{\rm bb}$ and $R_{\rm bb}$, for each source, we use the Markov chain Monte Carlo (MCMC) sampler {\tt emcee} \citep{Foreman-Mackey13}. We use 32 walkers and 5000 steps but discarding the first 2000 steps to ensure convergence. The best fitting results are presented in Table \ref{tab:fit}.

\begin{deluxetable*}{lcccccccccc}[!htbp]\label{tab:fit}
\tablecaption{Best fitting results for the blackbody model and maximum redshift}
\tablewidth{0pt}
\tablehead{
\ \ \ \,IAU  & $Swift$ & Fit & $T_{\rm bb}$ & $R_{\rm bb}$ & $L_{\rm bb}$ & $L_g$ & $z$ & $z_{\rm max}$ & $z_{\rm max}$ & $z_{\rm max}$\\
\ \ \,Name & UVOT & Peak & ($10^4$\,K) &  (log$_{10}$\,cm) & (log$_{10}$\,erg s$^{-1}$) & (log$_{10}$\,erg s$^{-1}$) & & ($r$=19.5) & ($r$=20.0) & ($r$=20.5)
}
\startdata
AT2018zr&Y&N&1.4&15.3&43.9&43.6&0.075&0.197&0.250&0.317\\
AT2018bsi&Y&N&2.3&14.8&43.9&43.1&0.051&0.104&0.133&0.170\\
AT2018hco&Y&N&2.3&14.9&44.1&43.4&0.088&0.141&0.180&0.231\\
AT2018hyz&Y&Y&1.8&15.1&44.1&43.6&0.046&0.192&0.245&0.314\\
AT2018iih&N&Y&3.4&15.1&45.2&44.0&0.212&0.314&0.413&0.551\\
AT2018jbv&Y&N&3.2&15.2&45.3&44.2&0.340&0.412&0.547&0.739\\
AT2018lni&N&Y&3.8&14.8&44.8&43.5&0.138&0.138&0.169&0.218\\
AT2018lna&Y&N&3.5&14.7&44.4&43.2&0.091&0.114&0.146&0.187\\
AT2019bhf&N&Y&2.5&14.9&44.3&43.4&0.121&0.158&0.202&0.261\\
AT2019cho&N&Y&2.4&15.0&44.3&43.5&0.193&0.193&0.216&0.279\\
AT2019azh&Y&Y&2.7&14.9&44.3&43.4&0.022&0.142&0.182&0.234\\
AT2019dsg&Y&Y&2.8&14.7&44.0&43.1&0.051&0.093&0.118&0.151\\
AT2019ehz&Y&Y&2.1&14.9&43.9&43.2&0.074&0.122&0.155&0.198\\
AT2019lwu&N&Y&1.7&15.1&43.8&43.4&0.117&0.145&0.185&0.236\\
AT2019meg&Y&N&2.5&14.9&44.3&43.4&0.152&0.154&0.197&0.254\\
AT2019mha&N&Y&1.6&15.1&43.8&43.3&0.148&0.148&0.184&0.234\\
AT2019qiz&Y&Y&1.8&14.8&43.5&42.9&0.015&0.079&0.100&0.128\\
AT2019teq&N&Y&1.2&15.0&43.3&43.0&0.088&0.097&0.123&0.155\\
AT2019vcb&N&Y&1.3&15.1&43.6&43.3&0.089&0.143&0.181&0.229\\
AT2020pj&N&Y&1.1&15.0&43.2&42.9&0.068&0.095&0.119&0.151\\
AT2020ddv&N&Y&3.4&14.8&44.6&43.4&0.160&0.160&0.189&0.244\\
AT2020ocn&Y&N&3.9&14.3&43.9&42.5&0.070&0.070&0.070&0.083\\
AT2020mbq&N&Y&1.4&15.0&43.5&43.2&0.093&0.115&0.146&0.185\\
AT2020mot&Y&Y&2.2&14.9&43.9&43.2&0.070&0.112&0.142&0.182\\
AT2020neh&Y&Y&1.7&15.0&43.7&43.2&0.062&0.123&0.156&0.199\\
AT2020nov&Y&Y&1.4&15.3&44.0&43.6&0.084&0.185&0.234&0.297\\
AT2020opy&Y&Y&1.7&15.2&44.1&43.6&0.159&0.197&0.251&0.321\\
AT2020qhs&Y&N&2.7&15.3&45.1&44.2&0.345&0.401&0.529&0.705\\
AT2020riz&N&Y&4.4&15.1&45.6&44.2&0.435&0.435&0.496&0.676\\
AT2020wey&Y&Y&2.6&14.4&43.3&42.4&0.027&0.046&0.058&0.074\\
AT2020vwl&Y&N&2.2&14.7&43.6&42.9&0.035&0.077&0.097&0.124\\
AT2020ysg&N&N&4.7&15.1&45.6&44.1&0.277&0.386&0.517&0.709\\
AT2020zso&Y&Y&1.8&15.0&43.9&43.3&0.061&0.128&0.163&0.208
\enddata    
\tablecomments{Explanations for the columns: \\1. $Swift$ UVOT: If the blackbody model fitting uses any reliable $Swift$ UVOT differential photometry (Y = yes, N = no).\\
2. Fit peak: If the blackbody model fitting is based on the photometry around the peak. Except AT2020ysg, all peaks are resolved, but some are either better sampled or followed by $Swift$ UVOT $\sim10-80$ days later, therefore we use the late-time observations to fit the blackbody temperature, then fix the temperature and fit the data around the peak to get the radius. Yes (Y) for fitting on the peak, no (N) for fitting on late-time observations.\\
3. $L_g$: The rest-frame $g$-band luminosity.\\
4. $z_{\rm max}$: The maximum redshift where this TDF could have been detected given the ZTF $r$-band effective limiting magnitude. It should not be less than $z$.}
\end{deluxetable*}

\section{Luminosity Function} \label{sec:lf}

After obtaining the peak blackbody parameters for each TDF in the ZTF-I sample, we can construct a peak LF with them. For a survey of sources with a constant flux, the LF can be estimated by weighting each source by the maximum volume, $V_{\rm max}$, in which the source can be detected \citep{Schmidt68}. For transients like TDFs, we are interested in their volumetric rate, i.e., the number of this kind of transients per cubic Mpc per year as a function of peak luminosity. 

\subsection{ZTF-I TDF LF} \label{sec:lfztf}
We calculate the $V_{\rm max}$ as follows. First, we determine the limiting magnitude. Since the ZTF-I TDFs are selected by $g-r<0$, the detection is limited by the $r$-band magnitude. The distribution of $r$-band magnitude indicates the limiting magnitude for $r$-band should be around $19.5\lesssim r\lesssim 20.5$. We set three limiting magnitudes: $r=19.5$, 20.0 and 20.5. For each limiting magnitude, we use the blackbody model parameters to calculate the maximum redshift $z_{\rm max}$ for each source. Meanwhile, we calculate the rest-frame peak luminosity in $g$ band, $L_g$. We show the $z_{\rm max}$ and $L_g$ in Table \ref{tab:fit}. 

After that, we set the survey area. The distribution of right ascension (R.A.) and declination (Dec.) for 33 TDFs yields a survey area of $A\approx15000$ deg$^2$. Next, the survey duration is set as $\tau=2.7$ yr. Finally, we get the volumetric rate $\dot{N}$ for each source.

Now we can build the luminosity function. We look into two kinds of luminosity: one is the rest-frame $g$-band peak luminosity, $L_g$, while the other one is the blackbody luminosity, $L_{\rm bb}$.

\subsubsection{LF for $L_g$}

For $L_g$, we first bin them into eight bins separated by equivalent log $L_g$, then sum up the $\dot{N}$ for each bin. The error for volumetric rate in each bin is estimated based on bootstrapping.

We use a power-law model to fit the volumetric rate as a function of $L_g$:
\begin{equation}
    \frac{d\dot{N}}{d\ {\rm log}_{10}L_g}=\dot{N}_0\left(\frac{L_g}{L_{g,0}}\right)^a.
\end{equation}
For $L_{g,0}=10^{43}$ erg s$^{-1}$, a least-square fit yields:
\begin{enumerate}
    \item For $r=19.5$, $\dot{N}_0=(7.8\pm2.2)\times10^{-8}$ Mpc$^{-3}$ yr$^{-1}$, $a=-1.28\pm0.22$.
    \item For $r=20.0$, $\dot{N}_0=(4.1\pm1.1)\times10^{-8}$ Mpc$^{-3}$ yr$^{-1}$, $a=-1.30\pm0.21$.
    \item For $r=20.5$, $\dot{N}_0=(2.1\pm0.6)\times10^{-8}$ Mpc$^{-3}$ yr$^{-1}$, $a=-1.33\pm0.20$.
\end{enumerate}

We notice the curves resemble a Schechter function \citep{Schechter76}, so we also fit each curve into a Schechter-like function, defined as
\begin{equation}\label{eqn:sch}
    \frac{d\dot{N}}{d\ {\rm log}_{10}L_g}=\dot{N}_0^*\left(\frac{L_g}{L_g^*}\right)^\alpha{\rm exp}\left(-\frac{L_g}{L_g^*}\right).
\end{equation}

The fitting results are as follows: 
\begin{enumerate}
    \item For $r=19.5$, ${\rm log}\,\dot{N}_0^*\,$(Mpc$^{-3}\,$yr$^{-1})=-7.62\pm0.23$, ${\rm log}\,L_g^*\,$(erg$\,$s$^{-1}$)$\,=43.82\pm0.14$, $\alpha=-0.76\pm0.11$. 
    \item For $r=20.0$, ${\rm log}\,\dot{N}_0^*\,$(Mpc$^{-3}\,$yr$^{-1})=-8.01\pm0.26$, ${\rm log}\,L_g^*\,$(erg$\,$s$^{-1}$)$\,=43.87\pm0.15$, $\alpha=-0.85\pm0.11$. 
    \item For $r=20.5$, ${\rm log}\,\dot{N}_0^*\,$(Mpc$^{-3}\,$yr$^{-1})=-8.34\pm0.27$, ${\rm log}\,L_g^*\,$(erg$\,$s$^{-1}$)$\,=43.88\pm0.16$, $\alpha=-0.89\pm0.11$. 
\end{enumerate}

The results are plotted in the left panel of Figure \ref{fig:lf}.

\subsubsection{LF for $L_{bb}$}

For $L_{\rm bb}$, we first bin them into seven bins separated by equivalent log $L_{\rm bb}$, then sum up the $\dot{N}$ for each bin. The error for volumetric rate in each bin is estimated based on bootstrapping.

We use a power-law model to fit the volumetric rate as a function of $L_{\rm bb}$:
\begin{equation}
    \frac{d\dot{N}}{d\ {\rm log}_{10}L_{\rm bb}}=\dot{N}_0\left(\frac{L_{\rm bb}}{L_{{\rm bb},0}}\right)^a.
\end{equation}
For $L_{{\rm bb},0}=10^{44}$ erg s$^{-1}$, a least-square fit yields:
\begin{enumerate}
    \item For $r=19.5$, $\dot{N}_0=(3.0\pm0.9)\times10^{-8}$ Mpc$^{-3}$ yr$^{-1}$, $a=-1.22\pm0.17$.
    \item For $r=20.0$, $\dot{N}_0=(1.6\pm0.5)\times10^{-8}$ Mpc$^{-3}$ yr$^{-1}$, $a=-1.23\pm0.17$.
    \item For $r=20.5$, $\dot{N}_0=(8.2\pm2.6)\times10^{-9}$ Mpc$^{-3}$ yr$^{-1}$, $a=-1.26\pm0.16$.
\end{enumerate}

\begin{figure*}[!htbp]
\plottwo{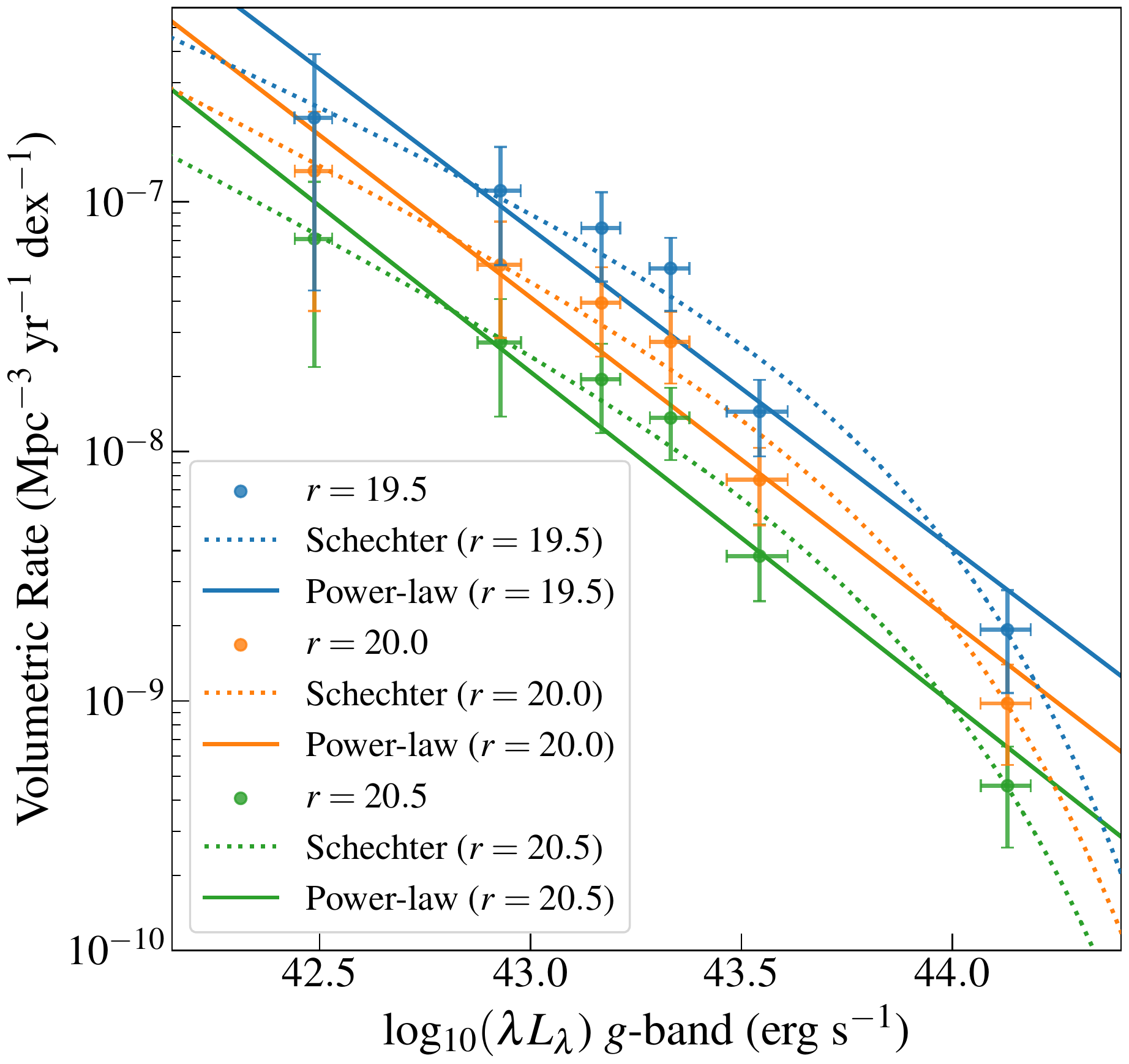}{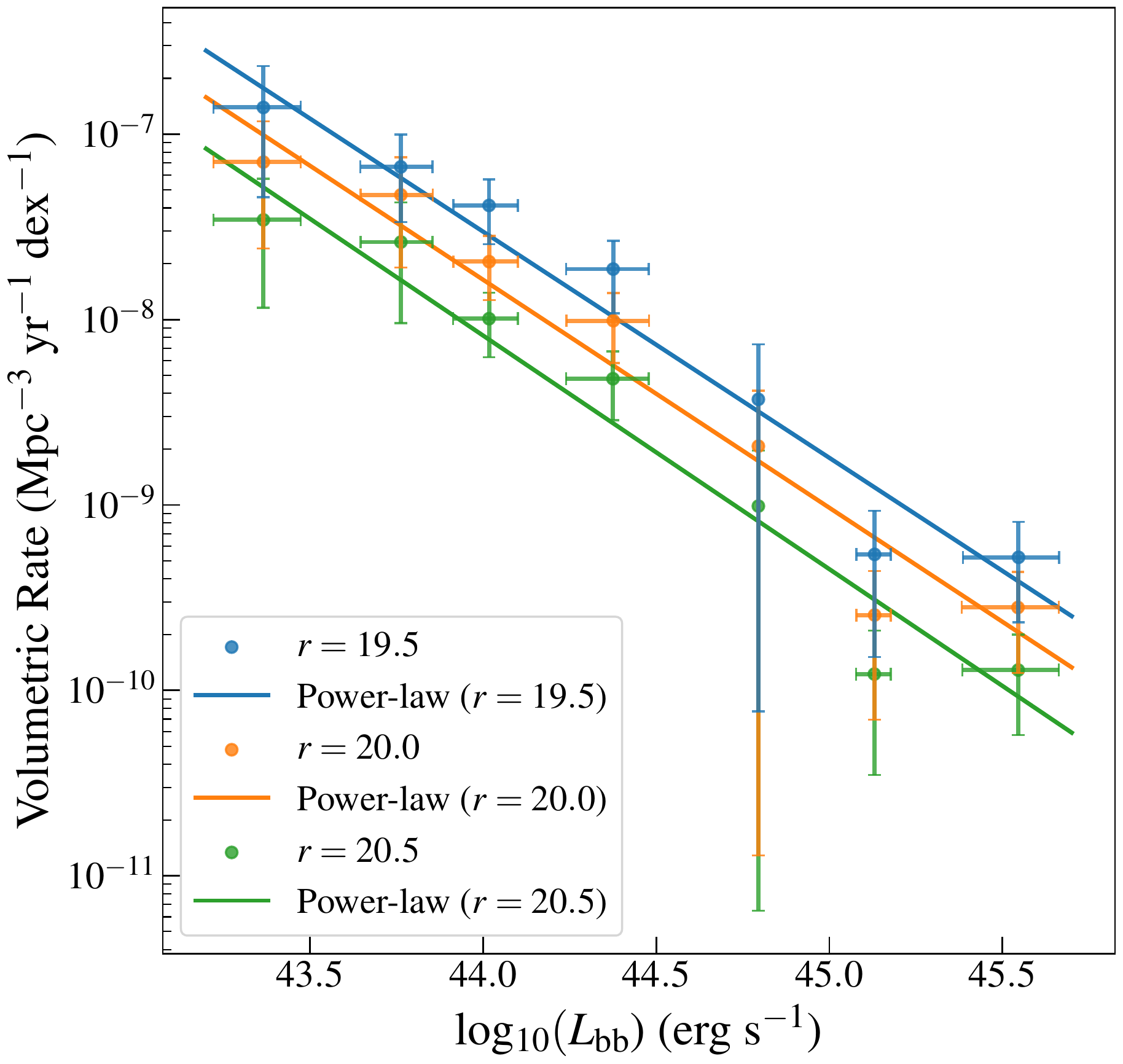}
\caption{The TDF LF based on 33 ZTF-I TDFs. The volumetric rates for $r$-band limiting magnitudes $r=19.5$, $20.0$ and $20.5$ are shown in blue, orange and green, respectively. The solid and dotted lines represent the fitting results to a power-law profile and a Schechter function, respectively. The error for volumetric rate in each bin is estimated based on bootstrapping. \textbf{Left:} LF for rest-frame $g$-band peak luminosity $L_g$. The sources are binned into eight bins separated by equivalent log $L_g$. The number of sources in these eight bins is \{2, 0, 4, 6, 8, 8, 0, 5\}.  \textbf{Right:} LF for blackbody luminosity $L_{\rm bb}$. The sources are binned into seven bins separated by equivalent log $L_{\rm bb}$. The number of sources in these seven bins is \{5, 7, 9, 6, 1, 2, 3\}.
\label{fig:lf}}
\end{figure*}

The results are plotted in the right panel of Figure \ref{fig:lf}.


\subsection{Comparison with Previous LF} \label{sec:vv}

Our procedures for constructing LFs are not identical with the previous LF. For comparison, we follow the procedures of \citet{vanVelzen18} to construct another set of LFs. We adopt the "1/$\mathcal{V}_{\rm max}$" method, in which $\mathcal{V}_{\rm max}$ is defined as
\begin{equation} \label{eqn:vmaxvv}
    \mathcal{V}_{\rm max}\equiv V(z_{\rm max})\ A_{\rm survey}\times\tau_{\rm survey}.
\end{equation}
In this equation, $V(z_{\rm max})$ represents the volume per unit solid angle corresponding to the maximum redshift, while $A_{\rm survey}\times\tau_{\rm survey}$ denotes the product of the effective survey duration and survey area. To estimate the maximum redshift, the flux limit or limiting magnitude for ZTF must be decided. We use the same set of limiting magnitudes: $r=19.5$, 20.0 and 20.5. After that, we define a typical TDF as a flare with a peak luminosity of $L_g^*=10^{42.5}$ erg s$^{-1}$ and temperature $T_{\rm bb}^*=2.5\times 10^4$ K, so that we can use the formula
\begin{equation} \label{eqn:ntdf}
    N_{\rm TDF, detected}\approx \dot{N}^*\times V(z_{\rm max}^*)\ A_{\rm survey}\times\tau_{\rm survey}
\end{equation}
to estimate the product of the effective survey duration and survey area $A_{\rm survey}\times\tau_{\rm survey}$. In this formula, the detected TDF number is $N_{\rm TDF, detected}=33$, while the mean volumetric rate $\dot{N}^*$ is set as $5\times10^{-7}$ Mpc$^{-3}$ yr$^{-1}$ following \citet{vanVelzen18}. The $z_{\rm max}^*$ for $r=19.5$, $r=20.0$ and $r=20.5$ are 0.052, 0.066 and 0.084, respectively. 

We use the blackbody model parameters and the limiting magnitudes to calculate $L_g$ and $z_{\rm max}$ for each source. The results are presented in Table \ref{tab:fit}. Next, we calculate $1/\mathcal{V}_{\rm max}$ for each source. We again look into two kinds of luminosity: $L_g$ and $L_{\rm bb}$. However, \citet{vanVelzen18} only provides the LF for $L_g$. Therefore, we first rebuild this LF, then use the blackbody parameters provided in Table 1 of \citet{vanVelzen18} to build the LF for $L_{\rm bb}$.

\subsubsection{LF for $L_g$}
For $L_g$, we bin all 33 ZTF-I TDFs into eight bins separated by equivalent log $L_g$, and sum up all $1/\mathcal{V}_{\rm max}$ in each bin. The uncertainties are estimated by $\sqrt{\sum 1/\mathcal{V}_{\rm max}^2}$. For $r=19.5$, the sum of $1/\mathcal{V}_{\rm max}$ for all 33 TDFs yields a rate of $6.3\times10^{-8}$ Mpc$^{-3}$ yr$^{-1}$, one order of magnitude lower than the rate based on 13 TDFs, $8\times10^{-7}$ Mpc$^{-3}$ yr$^{-1}$. In the left panel of Figure \ref{fig:lfvv}, we plot our result and the result of \citet{vanVelzen18} for comparison. The result indicates no correlation between the volumetric rate and the limiting magnitude, i.e., the volumetric rates for $r=19.5$, 20.0 and 20.5 are almost the same. According to \citet{vanVelzen20}, the single-epoch depth and filters of ZTF are similar to iPTF. Hence we adopt $r=19.5$, which is the effective limiting magnitude for iPTF in \citet{vanVelzen18}, for the following model fitting.

We use a power-law model to fit the volumetric rate as a function of $L_g$:
\begin{equation}
    \frac{d\dot{N}}{d\ {\rm log}_{10}L_g}=\dot{N}_0\left(\frac{L_g}{L_{g,0}}\right)^a.
\end{equation}
For $L_{g,0}=10^{43}$ erg s$^{-1}$, a least-square fit yields $\dot{N}_0=(5.1\pm1.5)\times10^{-8}$ Mpc$^{-3}$ yr$^{-1}$, $a=-1.28\pm0.22$.

For comparison, \citet{vanVelzen18} provided the fitting results. Notably, ASASSN-15lh is an exceptionally bright source among these 13 TDFs, superluminous yet with debating nature \citep[e.g.,][]{Dong16,Dai16,Leloudas16}. Therefore, model fitting parameters for samples with and without ASASSN-15lh are both provided. For samples with ASASSN-15lh, $a=-1.6\pm0.2$, $\dot{N}_0=(1.9\pm0.7)\times10^{-7}$ Mpc$^{-3}$ yr$^{-1}$. While for samples without ASASSN-15lh, $a=-1.3\pm0.3$, $\dot{N}_0=(2.3\pm0.8)\times10^{-7}$ Mpc$^{-3}$ yr$^{-1}$. Although the index $a$ is well consistent with the previous LF without ASASSN-15lh, it is higher than the previous LF with ASASSN-15lh, and the average of $\dot{N}_0$ is a factor of $\sim$5 lower than both previous LFs. Nonetheless, these differences are at $\sim2\sigma$ level. 

Alternatively, a Gaussian model 
\begin{equation}
    \frac{d\dot{N}}{d\ {\rm log}_{10}L_g}=\dot{N}_0'\ {\rm exp}\left\{-\frac{{[\rm log}_{10}(L_g/L_{g,0}')]^2}{2b^2}\right\}
\end{equation}
with $\dot{N}_0'=1.0\times10^{-7}$ Mpc$^{-3}$ yr$^{-1}$, $L_{g,0}'=10^{42.5}$ erg s$^{-1}$ and $b=0.54$ provides a more reasonable description for the LF. For comparison, \citet{vanVelzen18} used $\dot{N}_0'=1.0\times10^{-6}$ Mpc$^{-3}$ yr$^{-1}$, $L_{g,0}'=10^{42.5}$ erg s$^{-1}$ and $b=0.4$ to describe the LF for TDFs without ASASSN-15lh.

In the left panel of Figure \ref{fig:lfvv} we show the results for $r=19.5$, 20.0, 20.5 and \citet{vanVelzen18}. We will discuss these features in Section \ref{sec:dis}.

\subsubsection{LF for $L_{bb}$}
\begin{figure*}[!htbp]
\plottwo{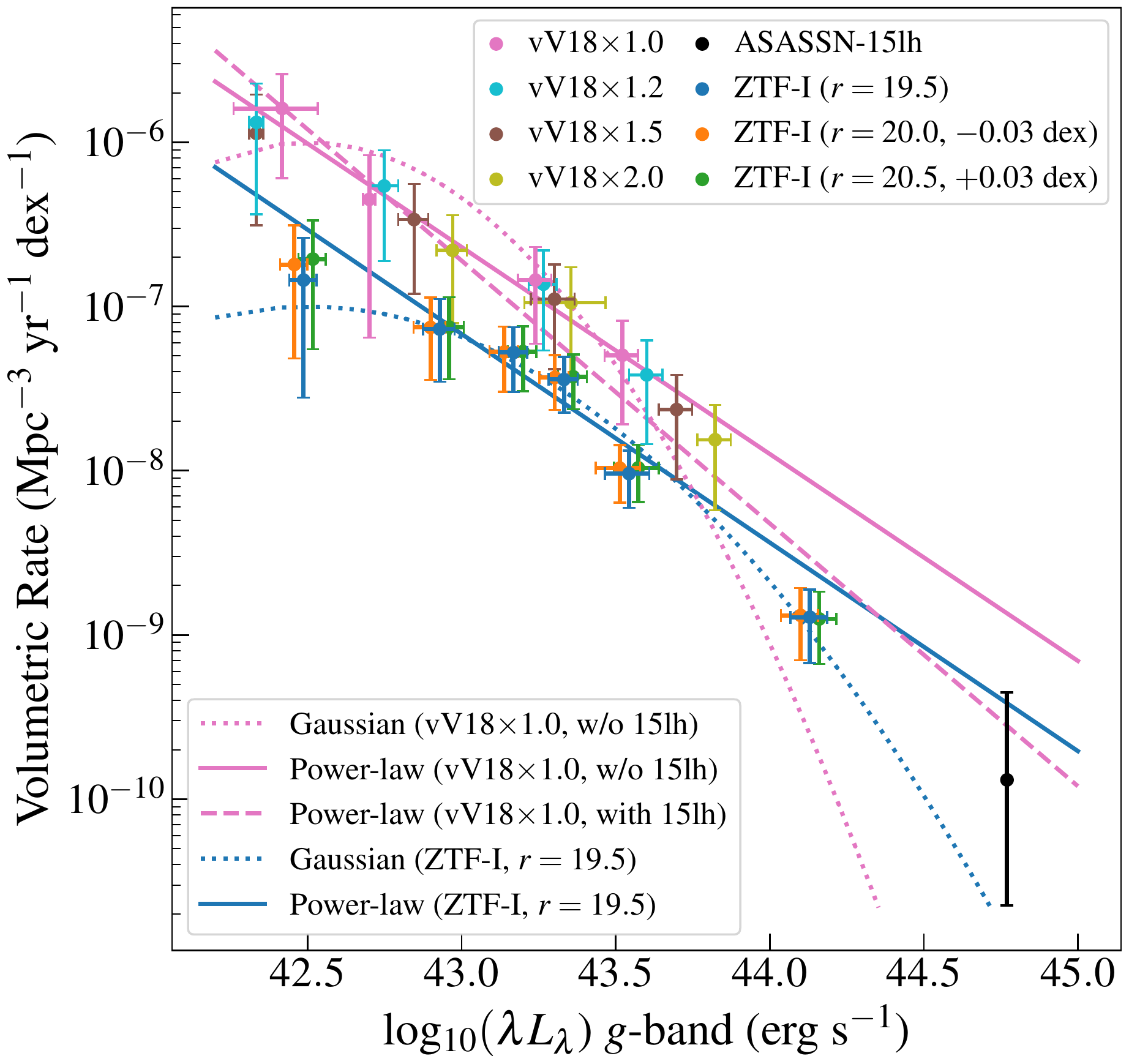}{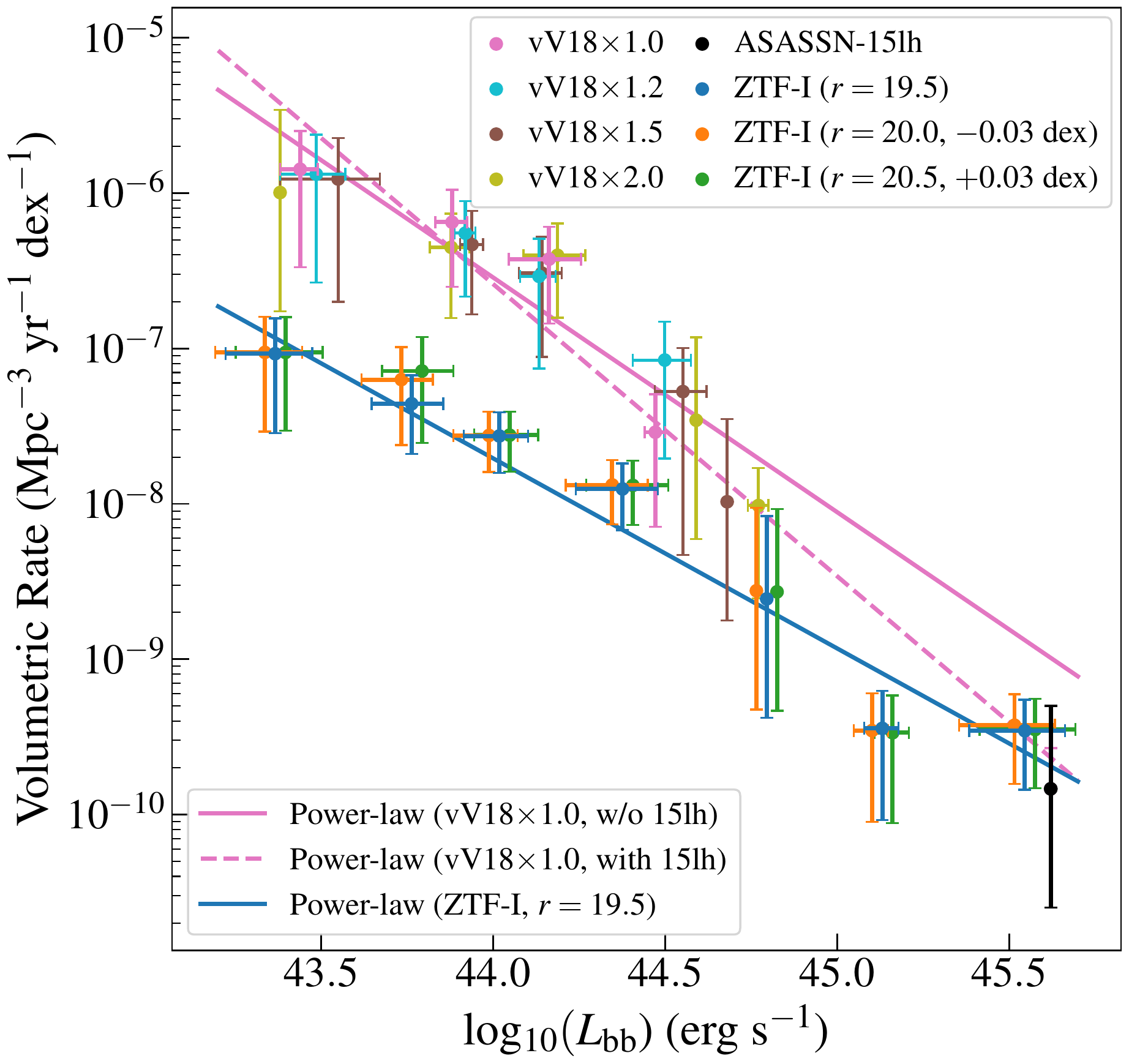}
\caption{The TDF LF based on 33 ZTF-I TDFs following the procedures of \citet{vanVelzen18}. The volumetric rates for limiting magnitudes $r=19.5$, 20.0 and 20.5 are shown in blue, orange and green, respectively. The orange and green markers are slightly shifted for the convenience of display. The blue solid and dashed represents the fitting result to a power-law and a Gaussian model for $r=19.5$, respectively. For comparison, the previous TDF LF by \citet{vanVelzen18} based on 13 TDFs is displayed in pink. Remarkably, the brightest source among 13 TDFs, ASASSN-15lh, is marked in black. It is outstandingly bright while its origin remains unclear. Therefore, we adopt the fitting parameters for both with and without (w/o) this source.\\
As 7 out of 13 TDFs in \citet{vanVelzen18} have only post-peak light curves, their $z_{\rm max}$ should be underestimated. Consequently, their volumetric rates are overestimated. To evaluate this effect, we multiply their peak luminosity for a factor of 1.2, 1.5, 2.0. The data points and error bars in pink, cyan, brown and olive correspond to multiplication factors ($\times$) of 1.0, 1.2, 1.5 and 2.0, respectively.
\textbf{Left:} LF for rest-frame $g$-band peak luminosity $L_g$. The sources are binned into eight bins separated by equivalent log $L_g$. The number of sources in these eight bins is \{2, 0, 4, 6, 8, 8, 0, 5\}. While for the sources of \citet{vanVelzen18}, the sources are binned into seven bins separated by equivalent log $L_g$. The number of sources in these seven bins for the multiplication factor of 1.0 (pink) is \{4, 2, 3, 3, 0, 0, 1\}. When the multiplication factors for post-peak sources are applied, the original power-law profile (solid and dashed pink line) can still describe the LF, while the Gaussian profile (dashed pink line) can not.
\textbf{Right:} LF for blackbody luminosity $L_{\rm bb}$. The sources are binned into seven bins separated by equivalent log $L_{\rm bb}$. The number of sources in these seven bins is \{5, 7, 9, 6, 1, 2, 3\}. While for the sources of \citet{vanVelzen18}, the sources also are binned into seven bins separated by equivalent log $L_{\rm bb}$. The number of sources in these seven bins for the multiplication factor of 1.0 (pink) is \{2, 4, 4, 2, 0, 0, 1\}. When the multiplication factors for post-peak sources are applied, the original power-law profile (solid and dashed pink line) can still describe the LF.
\label{fig:lfvv}}
\end{figure*}

For $L_{\rm bb}$, we bin all 33 ZTF-I TDFs into seven bins separated by equivalent log $L_{\rm bb}$, and sum up all $1/\mathcal{V}_{\rm max}$ in each bin. The uncertainties are estimated by $\sqrt{\sum 1/\mathcal{V}_{\rm max}^2}$. For comparison, we use the blackbody parameters provided in Table 1 of \citet{vanVelzen18} to build the LF for $L_{\rm bb}$. In the right panel of Figure \ref{fig:lfvv}, we plot the LFs for $L_{\rm bb}$ of ZTF-I and \citet{vanVelzen18} TDFs. The ZTF-I TDF LFs again indicate no correlation between the volumetric rate and the limiting magnitude, hence we adopt $r=19.5$ for the following model fitting.

We use a power-law model to fit the volumetric rate as a function of $L_{\rm bb}$:
\begin{equation}
    \frac{d\dot{N}}{d\ {\rm log}_{10}L_{\rm bb}}=\dot{N}_0\left(\frac{L_{\rm bb}}{L_{\rm{bb,0}}}\right)^a.
\end{equation}
For $L_{\rm{bb,0}}=10^{44}$ erg s$^{-1}$, a least-square fit yields:
\begin{enumerate}
    \item For ZTF-I TDFs, $\dot{N}_0=(1.9\pm0.6)\times10^{-8}$ Mpc$^{-3}$ yr$^{-1}$, $a=-1.22\pm0.18$. 
    \item For \citet{vanVelzen18} TDFs with ASASSN-15lh, $\dot{N}_0=(2.5\pm1.0)\times10^{-7}$ Mpc$^{-3}$ yr$^{-1}$, $a=-1.89\pm0.21$. 
    \item For \citet{vanVelzen18} TDFs without ASASSN-15lh, $\dot{N}_0=(2.8\pm1.1)\times10^{-7}$ Mpc$^{-3}$ yr$^{-1}$, $a=-1.51\pm0.42$.
\end{enumerate}

The LF of ZTF-I TDFs is much shallower than that of \citet{vanVelzen18} TDFs. In the right panel of Figure \ref{fig:lfvv} we show the results for $r=19.5$, 20.0, 20.5 and \citet{vanVelzen18}. We will discuss these features in Section \ref{sec:dis}.

\section{Discussions} \label{sec:dis}
\subsection{Shape of the LF}
First, we discuss the shape of the LF.
The ZTF-I TDF LF for $L_g$ can be described by a power-law profile, $dN/dL_g\propto L_g^{-2.3\pm0.2}$. This is consistent with the previous \citet{vanVelzen18} LF ($dN/dL_g\propto L_g^{-2.6\pm0.2}$ for all 13 TDFs, $dN/dL_g\propto L_g^{-2.3\pm0.3}$ if ASASSN-15lh is excluded).
Additionally, it can be well described by a Schechter-like function. While for $L_{\rm bb}$, it can also be described by a power-law profile, $dN/dL_{\rm bb}\propto L_{\rm bb}^{-2.2\pm0.2}$. This is shallower than the \citet{vanVelzen18} LF ($dN/dL_{\rm bb}\propto L_{\rm bb}^{-2.9\pm0.2}$ for all 13 TDFs, $dN/dL_{\rm bb}\propto L_{\rm bb}^{-2.5\pm0.4}$ if ASASSN-15lh is excluded).


At the low-luminosity end, the LF for $L_g$ flattens at $L_g\sim10^{42.5-43.0}$ erg s$^{-1}$. Given a bolometric correction of $\sim10$, it corresponds to the Eddington luminosity for BHs with mass $M_{\rm BH}=10^{5.5-6.0}\ M_{\odot}$. This is consistent with an Eddington-limited emission scenario. However, a power-law profile can still fit this end due to the large errors, and the LF for $L_{\rm bb}$ does not flatten at low luminosity. Hence, more faint samples are needed to reach a final judgment.

While at the high-luminosity end, 5 bright sources constrain the volumetric rate at $L_g\sim 10^{44.0-44.3}$ erg s$^{-1}$ and $L_{\rm bb}\sim 10^{45.0-45.7}$ erg s$^{-1}$, indicating a suppression of volumetric rate in this range.
This corresponds to the Eddington limit for BHs with $M_{\rm BH}\gtrsim10^{7}\ M_{\odot}$.  The TDF can only be created when the tidal radius ($R_{\rm t}$) equals or surpasses the innermost bound circular orbit (IBCO) of the black hole, i.e., $R_{\rm t}\gtrsim R_{\rm IBCO}$, indicating a maximum black hole mass, or the Hills mass. For non-spinning BHs, 
\begin{equation}
    M_{\rm Hills}=9.0\times10^7\,M_{\odot}\left(\frac{M_{\star}}{M_{\odot}}
    \right)^{-1/2}\left(\frac{R_{\star}}{R_{\odot}}\right)^{3/2}
\end{equation}
\citep{Hills75,Beloborodov92,Leloudas16}.
This indicates a rate suppression around $L_g\sim10^{45}$ erg s$^{-1}$ and $L_{\rm bb}\sim10^{46}$ erg s$^{-1}$ for sun-like stars. However, the vast majority of stars should be sub-solar main sequence stars, which obey $M_{\rm Hills}\propto M_{\star}^{0.7}$ \citep{Stone16,Lin22}. Therefore, the average Hills mass should be lower, yielding a more consistent, lower luminosity. The above result agrees with the recent calculation of \citet{Coughlin22}, which indicates a rate suppression for BHs with $M_{\rm BH}>10^{7}\ M_{\odot}$, given a predominantly low-mass stellar population.

\subsection{Volumetric Rate}

As introduced in Section \ref{sec:vv}, the sum of $1/\mathcal{V}_{\rm max}$ for all 33 TDFs yields a rate of $6.3\times10^{-8}$ Mpc$^{-3}$ yr$^{-1}$, that is one order of magnitude lower than the rate inferred from the 13 TDFs in \citet{vanVelzen18}, $8\times10^{-7}$ Mpc$^{-3}$ yr$^{-1}$. Note that 7 out of 13 TDFs have only post-peak light curves, which will certainly lead to an overestimate of the volumetric rate. In order to assess the impact of these post-peak TDFs, 
for each source that only has post-peak data, we multiply the $L_g$ and $L_{\rm bb}$ by factors of 1.2, 1.5 and 2.0, respectively. Then we recalculate new $z_{\rm max}$ and the $\mathcal{V}_{\rm max}$ for each multiplication factor. Finally, we add up the $1/\mathcal{V}_{\rm max}$ and get new LFs. The results are shown in Figure \ref{fig:lfvv}.


As displayed in Figure \ref{fig:lfvv}, if the true peak luminosity of all of the sources discovered after maximum light is higher than the observed maximum luminosity, the original Gaussian profile will not stand. However, the original power-law profile seems stable within the 1-$\sigma$ errors, while a systematically higher volumetric rate than the ZTF-I TDFs still remains. The sum of $1/\mathcal{V}_{\rm max}$ for all 13 TDFs that used for the LF yields a rate of (7.2, 6.6, 6.0) $\times10^{-7}$ Mpc$^{-3}$ yr$^{-1}$ for factors of 1.2, 1.5 and 2.0. Therefore, we conclude that the sources detected after the peak only have limited effect on the volumetric rate. 

We notice that the 13 sources are collected from 5 surveys ranging from 2004 to 2016, a normalization based on the TDF number is applied (Equation \ref{eqn:ntdf}), the product of effective survey duration and survey area ($A\times\tau$) ranges from 17 deg$^2$ yr ($GALEX$) to 82637 deg$^2$ yr (ASAS-SN). As a result, the serendipity of the TDF discoveries may lead to a higher (or lower) volumetric rate. For example, the two SDSS TDFs contribute $1.3\times10^{-7}$ and $8.6\times10^{-9}$ Mpc$^{-3}$ yr$^{-1}$ to the total volumetric rate, while the two iPTF sources contribute $6.3\times10^{-9}$ and $3.2\times10^{-7}$ Mpc$^{-3}$ yr$^{-1}$ to the total volumetric rate. Limited by the TDF number of each survey, we can not decide whether this serendipity raises or suppresses the volumetric rate. The reason for the lower volumetric rate is still uncertain.

\section{Conclusions} \label{sec:con}

We have obtained the optical and blackbody LFs for the 33 TDFs discovered in the ZTF-I survey, which is the largest sample obtained in a systematic search up to now. In addition to a much larger sample size, majority of these TDFs have nice coverage around their luminosity peaks and have been continuously followed by $Swift$ observations. We have carefully calculated the rest-frame $g$-band luminosity $L_g$ and blackbody luminosity $L_{\rm bb}$ by blackbody fitting to the SEDs around peaks and then the LFs. For comparison, we also use the data of \citet{vanVelzen18} sample to rebuild the optical LF, and build a blackbody LF. Our conclusions are summarized as follows.

\begin{enumerate}
\item The LF for $L_g$ can be described by a power-law profile, $dN/dL_g\propto L_g^{-2.3}$. This is consistent with the previous LF by \citet{vanVelzen18}, albeit the normalization factor $\dot{N}_0$ is $\sim5$ times lower. It can also be well described by a Schechter-like function (Equation \ref{eqn:sch}).
\item The LF for $L_{\rm bb}$ can be described by a power-law profile, $dN/dL_{\rm bb}\propto L_{\rm bb}^{-2.2}$, which is shallower than the LF of the \citet{vanVelzen18} sample.
\item At the low-luminosity end, the flat profile at $L_g\sim10^{42.5-43.0}$~erg s$^{-1}$ supports an Eddington-limited emission mechanism, as the luminosity corresponds to the Eddington luminosity for BHs with mass $M_{\rm BH}=10^{5.5-6.0}\ M_{\odot}$, given a bolometric correction of $\sim10$. However, the blackbody LF does not show a corresponding flat profile.
\item At the high-luminosity end, the volumetric rate drops at $L_g>10^{44}$ erg s$^{-1}$ and $L_{\rm bb}>10^{45}$ erg s$^{-1}$, corresponding to the Eddington luminosity for BHs with $M_{\rm BH}\gtrsim10^{7}\ M_{\odot}$. This is consistent with a rate suppression around the Hills mass.
\item The total volumetric rate is one order of magnitude lower than that given by \citet{vanVelzen18}. Correcting the peak luminosity for sources observed post peaks in \citet{vanVelzen18} can not effectively eliminate the discrepancy. The previous LF construction might be yet greatly impacted by the serendipitous discoveries in the step of normalization of these surveys (Equation \ref{eqn:ntdf}). However, how the serendipity affects volumetric rate in detail remains unclear due to the small number of TDFs in each survey.
\end{enumerate}

The uniform, wide-field and high-cadence ZTF-I survey greatly benefits the systematical search for TDFs, and thus well reduces the serendipity and improves the reliability of the LFs. Upcoming similar but deeper surveys, for instance, by the Vera Rubin Observatory (VRO) and Wide-Field Survey Telescope (WFST), should unveil much more TDFs, especially further and fainter ones. Hence, they should pave the way to more accurate LFs, and better constrain the shape of the LFs, especially on the faint end, e.g., towards a power-law profile, a Schechter-like function, or other formats.

\begin{acknowledgments}
We thank the anonymous referee for constructive comments. This work is supported by the Strategic Priority Research Program of Chinese Academy of Sciences (No. XDB 41000000), the National Key R\&D Program of China (2017YFA0402600), the National
Science Foundation of China (NSFC) grants (No. 12233008, 11833007, 11973038, 12073025, 12192221), the Fundamental Research Funds for the Central Universities, and the China Manned Space Project (Nos. CMS-CSST-2021-A07, CMS-CSST-2021-B11). The authors also gratefully acknowledge the support of Cyrus Chun Ying Tang Foundations.
The ZTF forced-photometry service was funded under the Heising-Simons Foundation grant \#12540303 (PI: Graham).
\end{acknowledgments}






\bibliography{sample631}{}
\bibliographystyle{aasjournal}



\end{document}